\documentclass{epl}

\title{Particle-Based Mesoscale Hydrodynamic Techniques}
\shorttitle{Mesoscale hydrodynamic techniques}

\author{Hiroshi Noguchi\inst{1}\thanks{E-mail:\email{hi.noguchi@fz-juelich.de}} \and Norio Kikuchi\inst{2} \and Gerhard Gompper\inst{1} }
\shortauthor{H. Noguchi \etal}

\institute{
\inst{1} Institut f\"ur Festk\"orperforschung, Forschungszentrum J\"ulich, 
52425 J\"ulich, Germany\\
\inst{2} Fachbereich Physik, Martin-Luther-Universit\"at Halle-Wittenberg, D-06099 Halle, Germany
}


\begin{document}
\maketitle

\begin{abstract}
Dissipative particle dynamics (DPD) and multi-particle collision (MPC) 
dynamics are powerful tools to study mesoscale hydrodynamic phenomena
accompanied by thermal fluctuations.
To understand the advantages of these types of mesoscale simulation 
techniques in more detail, we propose new two methods, which are 
intermediate between DPD and MPC ---  
DPD with a multibody thermostat (DPD-MT), and MPC-Langevin dynamics (MPC-LD).
The key features are applying a Langevin thermostat to
the relative velocities of pairs of particles or multi-particle collisions,
and whether or not to employ collision cells.
The viscosity of MPC-LD is derived analytically, in very good agreement 
with the results of numerical simulations.
\end{abstract}

\begin{figure}
\onefigure{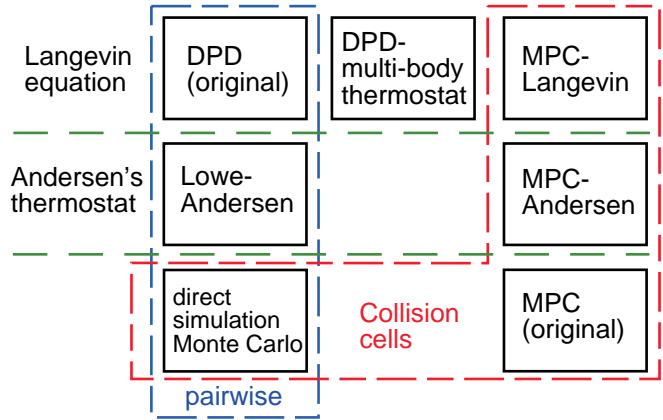}
\caption{  \label{fig:rela}
(Color online)
Relations between particle-based hydrodynamic methods.
}
\end{figure}

Soft matter systems such as polymer solutions, colloidal suspensions, vesicles, 
cells, and microemulsions exhibit many interesting dynamical 
behaviors, where hydrodynamic flow plays an important role, as do thermal 
fluctuations.
Several mesoscale simulation techniques for the flow of complex fluids 
accompanied by thermal fluctuations have been developed
in the last decades, such as direct simulation Monte Carlo (DSMC) 
\cite{bird76,bird98}, the Lattice Boltzmann method \cite{succ01},
dissipative particle dynamics (DPD) 
\cite{hoog92,groo97,bone97,espa97a,pago01,espa03,shar03,niku03,pete04},
and multi-particle collision (MPC) dynamics 
\cite{male99,ihle01,alla02,kiku03,ihle06}.
These methods have many similarities.  The most important common feature 
is the local mass and momentum conservation,
which is crucial to obtain hydrodynamic behavior in the continuum limit.
Since most of these methods were developed independently,
the relations between them are not well explored so far.
In this letter, we clarify the relations between the particle-based 
off-lattice methods,
particularly DPD and MPC, and use this insight to propose two new 
intermediate methods (all summarized in Fig.~\ref{fig:rela}).

We start from the relations between the Langevin and the Andersen's 
thermostat (AT) \cite{ande80}.
The underdamped Langevin equation of $N$ particles is given by
\begin{eqnarray} 
\label{eq:lgv}
m \frac{d {\bf v}_{i}}{dt} =
 - \nabla_i U + f_{\rm {LT}}, \\
f_{\rm {LT}} = - \gamma {\bf v}_{i} + \sigma{\bf \xi}_{i}(t),
\end{eqnarray} 
where $\nabla_i =\partial/\partial {\bf r}_i$,
$m$ is the mass of a fluid particle, and ${\bf r}_{i}$ and ${\bf v}_{i}$ 
are the position and velocity of the $i$-th particle, respectively.
The force $f_{\rm {LT}}$ represent the Langevin thermostat.
To satisfy the fluctuation-dissipation theorem,
the Gaussian white noise ${\bf \xi}_{i}(t)$ has to have  
the average $\langle \xi_{i,\alpha}(t) \rangle  = 0$ and the variance
$\langle \xi_{i,\alpha}(t) \xi_{j,\beta}(t')\rangle  =  
         2 k_{\rm B}T \delta _{ij} \delta _{\alpha\beta} \delta(t-t')$,
where $\alpha, \beta \in \{x,y,z\}$ and $k_{\rm B}T$ is the thermal energy, 
and its amplitude $\sigma$ in Eq.~(\ref{eq:lgv}) is related to the friction constant $\gamma$ by $\gamma=\sigma^2$. 
Each particle has one thermostat (heat bath), which is independent of all 
other particles.
This thermostat does not conserve momentum, and hence
the hydrodynamic interactions are not taken into account.
We separately integrate the potential forces 
$-\nabla_i U$ and the Langevin thermostat $f_{\rm {LT}}$ using a 
multiple-time-step algorithm \cite{tuck92}, where a shorter time step is 
employed for $-\nabla_i U$. The Langevin thermostat is integrated by the 
leapfrog algorithm \cite{alle87}, which implies
\begin{eqnarray}\label{eq:lgvlf}
{\bf v}_{i}(t_{n+1}) &=& a{\bf v}_{i}(t_n) + b{\bf \xi}_{i,n},\\
{\rm where\ } a &=& \frac{1-\gamma \Delta t/2m}{1+\gamma \Delta t/2m},\  
b= \frac{\sqrt{\gamma\Delta t}/m}{1+\gamma \Delta t/2m}
\label{eq:lgvab}
\end{eqnarray} 
with ${\bf v}_{i}(t_{n+1/2})= \{{\bf v}_{i}(t_{n+1})+{\bf v}_{i}(t_n)\}/2$
and ${\bf \xi}_{i}(t_{n+1/2})={\bf \xi}_{i,n}/\sqrt{\Delta t}$.
The modified Verlet algorithm in Ref.~\cite{shar03} also gives Eq.~(\ref{eq:lgvlf}).
This thermostat works even with large time steps $\Delta t > 2m/\gamma$.
For the thermodynamically ideal gas ($U=0$), Eq.~(\ref{eq:lgvlf}) gives
${\bf v}_{i}(t_n) = \sum_{l=0}^{\infty} a^l b{\bf \xi}_{i,n-l-1}$.
This velocity exhibits a Maxwell-Boltzmann distribution with 
$\langle v_{i,\alpha}(t_n)v_{i,\alpha}(t_n)\rangle = k_{\rm B}T/m$ for 
any $\Delta t$.
Thus, this discretized thermostat belongs to the class of generalized ATs 
described in Ref.~\cite{pete04}. For $\gamma \Delta t/2m=1$, the first 
term in Eq.~(\ref{eq:lgvlf}) vanishes ($a=0$),
and a new velocity ${\bf v}_{i}(t_{n+1})$ is selected from a 
Maxwell-Boltzmann distribution.
This corresponds to the original Andersen's thermostat \cite{ande80}.
Thus, AT can be interpreted as the discrete version of the Langevin 
thermostat with $\gamma \Delta t/2m=1$. 
This relation between the Langevin and Andersen's thermostat remains valid 
in DPD and MPC, as shown below.

The DPD thermostat is a modified Langevin thermostat,
which applies the relative velocities of the neighbor pairs. This implies
that the thermal force in Eq.~(\ref{eq:lgv}) is now given by 
\begin{equation}\label{eq:dpdorg}
 f_{\rm {DT0}}= \sum_{j\not=i} \left\{-w(r_{ij})({\bf v}_{i}-{\bf v}_{j} )
   \cdot{\bf \hat{r}}_{ij} + 
      \sqrt{w(r_{ij})}{\xi}_{ij}(t)\right\}{\bf \hat{r}}_{ij},
\end{equation} 
where ${\bf r}_{ij}= {\bf r}_{i}-{\bf r}_{j}$, ${r}_{ij}=|{\bf r}_{ij}|$, 
${\bf \hat{r}}_{ij}={\bf r}_{ij}/{r}_{ij}$, and ${\xi}_{ji}(t)=-{\xi}_{ij}(t)$.
This thermostat is applied only in the direction ${\bf \hat{r}}_{ij}$ 
to conserve the local angular momentum.
In DPD, a linear weight function, with 
$w(r_{ij})=\gamma(1-r_{ij}/r_{\rm {cut}})$ for $r_{ij}/r_{\rm {cut}}<1$ 
and $w(r_{ij})=0$ otherwise, is typically employed. Furthermore, 
DPD is usually combined with soft repulsive potentials 
$U(r_{ij})=(A/2)(1-r_{ij}/r_{\rm {cut}})^2$ 
with cutoff at $r_{\rm {cut}}$ \cite{groo97},
but other pairwise or multibody potentials are also available \cite{pago01}.
DPD shares many features with smooth-particle hydrodynamics 
(SPH)~\cite{mona92}, a method to solve the Navier-Stokes equation 
in a Lagrangian representation, and a modified version \cite{espa03} 
of DPD corresponds to SPH with thermal fluctuations.
In Shardlow's splitting algorithm~\cite{shar03}, each thermostat of the 
$ij$ pair is separately integrated.
This algorithm with multiple time steps gives 
the Andersen-thermostat version of DPD proposed by Lowe~\cite{lowe99}.
An energy-conservation version of DPD (micro-canonical ensemble), called DPD+e, has also 
been proposed~\cite{bone97,espa97a}.
DPD+e needs an additional variable in order to exchange the momenta of 
particles, since five of six degrees of freedom of a particle-pair are fixed
by the conservation of the translational ($3$) and angular ($2$) momenta.
We modify the DPD thermostat to remove angular-momentum conservation 
(DPD-a) for the discussions below. In this case, the thermal force
reads
\begin{equation}\label{eq:dpdnan}
 f_{\rm {DT1}}= \sum_{j\not=i} -w(r_{ij})({\bf v}_{i}-{\bf v}_{j} ) + \sqrt{w(r_{ij})}{\bf \xi}_{ij}(t).
\end{equation} 
We call the versions of methods with or without angular-momentum conservation '+a' or '-a', respectively.
$f_{\rm {DT1}}$ still keeps the translational momentum conservation.
Note that the numbers of the thermostats are $N N_{\rm {nb}}/2$ and $3N N_{\rm {nb}}/2$
for $f_{\rm {DT0}}$ and $f_{\rm {DT1}}$, respectively,
where $N_{\rm {nb}}$ is the mean number of the neighbors with $r_{ij}<r_{\rm {cut}}$.
These can be much more than the number of degrees of freedom $3N$.

MPC is a modification of DSMC to include multi-particle collision, 
in order to make the algorithm more efficient in its application to 
fluids \cite{male99}.
It is also called stochastic rotational dynamics (SRD) \cite{ihle01}.
The MPC algorithm consists of alternating streaming and collision steps. 
In the streaming step, the particles move ballistically, 
${\bf r}_{i}(t+\Delta t) 
    = {\bf r}_{i}(t) + {\bf v}_{i} \Delta t$,
where $\Delta t$ 
is the time interval between collisions. In the collision step, 
the particles are sorted into cubic cells of lattice constant $l_{\rm c}$.
The collision step consists of a stochastic rotation of the 
relative velocities of each particle in a cell, 
\begin{equation}\label{eq:mpc0}
{\bf v}_{i}^{\rm {new}}= {\bf v}_{\rm c}^{\rm G} + 
           {\bf \Omega}\{{\bf v}_{i}-{\bf v}_{\rm c}^{\rm G} \},
\end{equation}
where ${\bf v}_{\rm c}^{\rm G}$ is the velocity of the center of mass of all 
particles in the cell. The matrix ${\bf \Omega}$ rotates 
velocities by the angle $\phi$ around an axis, which is chosen 
randomly for each cell.
The translational-momentum and kinetic energy are conserved in the cell.
The collision cells are randomly shifted before each collision step to ensure  
Galilean invariance~\cite{ihle01}.
Although MPC fluid  originally corresponds to micro-canonical ensemble,
the temperature can be controlled by 
an additional rescaling of the relative velocities  ${\bf v}_{i}-{\bf v}_{\rm {c}}^{\rm G}$.
In MPC, the angular momentum is not conserved and the rotational symmetry 
is broken by the use of cells.
In DSMC~\cite{bird76}, two particles collide in the cell instead.
Thus, the difference between DSMC and MPC is whether collisions affect
two or all particles in the cell.

An Andersen-thermostat version of MPC (MPC-AT) has also been 
proposed~\cite{alla02}.  In MPC-AT, the velocities in the collision step are
obtained as
\begin{equation}\label{eq:mpcat}
{\bf v}_{i}^{\rm {new}}= {\bf v}_{\rm c}^{\rm G} + {\bf v}_{i}^{\rm {ran}} 
  - \sum_{j \in {\rm cell}} {\bf v}_j^{\rm {ran}}/N_{\rm {c}},
\end{equation}
where ${\bf v}_{i}^{\rm {ran}}$ is a velocity chosen from a 
Maxwell-Boltzmann distribution
and $N_{\rm {c}}$ is the number of particles in a cell.
The summation in Eq.~(\ref{eq:mpcat}) and the other equations in MPC 
runs over all particles in a cell.
The velocity of the center of mass of each cell is conserved, and
the temperature is constant in MPC-AT (instead of the energy in MPC).

\begin{table}
\caption{\label{tab:ande} 
Comparison between DPD and MPC methods
of the Langevin dynamics and Andersen-thermostat versions.}
\begin{center}
\begin{tabular}{lll}
\hline
               &  DPD  & MPC \\
 \hline
interacting particles: & & \\
\ \ number         &   $2$       &  multiple  \\
\ \ chosen by      &  distance $r_{ij}$  &  collision cell  \\
 & & \\
potential interaction  &  available       &  available   \\
momentum conservation:  & & \\
\ \ translation &  yes  & yes \\ 
\ \  angle &  on/off  & on/off \\ 
energy conservation   &  available   &  available    \\
\hline
\end{tabular}
\end{center}
\end{table}

In order to compare the various methods summarized in 
Table~\ref{tab:ande}, we have to distinguish the differences between
methods from the variations within each method.
Since both of DPD and MPC have Andersen-thermostat versions,
we compare them first.
Several differences originate from the variations of each method.
The MPC fluid is originally an ideal gas from a thermodynamic point
of view. It has been generalized by  Ihle {\it et al.} \cite{ihle06} 
by a modification of the collision rule to produce a nonideal-gas
equation of state. 
However, the equation of state can also be changed by the usual 
potential interactions, when the particles move according to 
Newton's equation $m d {\bf v}_{i}/dt = - \nabla_i U$ in the streaming step.
The angular momentum is not conserved in original MPC-AT.
However, it can be conserved by the addition of an angular-momentum 
constraint, such that
\begin{equation}\label{eq:mpcatan}
{\bf v}_{i}^{\rm {new}}= 
 {\bf v}_{\rm c}^{\rm G} + {\bf v}_{i}^{\rm {ran}}
 - \sum_{j \in {\rm cell}} {\bf v}_j^{\rm {ran}}/N_{\rm {c}}
 + \left\{ m{\bf \Pi}^{-1} \sum_{j \in {\rm cell}} {\bf r}_j\times 
     ({\bf v}_j-{\bf v}_j^{\rm {ran}})\right\}\times {\bf r}_{i},
\end{equation}
where ${\bf \Pi}$ is the moment-of-inertia tensor of the particles in the cell.
The energy can also be conserved in MPC-AT by the velocity scaling,
${\bf u}_{i}^{\rm {new}}\rightarrow \sqrt{\sum {u_{j}}^2/\sum(u_{j}^{\rm {new}})^2} 
{\bf u}_{i}^{\rm {new}}$,
where the relative velocity is ${\bf u}_{i}={\bf v}_{i}-{\bf v}_{\rm {c}}^{\rm G}$ 
(MPC-AT-a) or  
${\bf u}_{i}={\bf v}_{i}-{\bf v}_{\rm {c}}^{\rm G}-
\{m{\bf \Pi}^{-1} \sum {\bf r}_j\times {\bf v}_j\}\times {\bf r}_{i}$ 
(MPC-AT+a)
after the procedure of Eq.~(\ref{eq:mpcat}) or (\ref{eq:mpcatan}).
Note that an additional variable is {\em not} necessary in the energy-conserved 
MPC-AT, since undetermined degrees of freedom remain for $N_{\rm c}\ge3$.
Thus, two key features can be identified as the genuine differences between 
DPD and MPC: (i)  
The thermostats act on the relative velocity of two (DPD) or multiple (MPC) 
particles, and (ii)
the interacting particles are chosen by their distance $r_{ij}$ (DPD) or 
by sharing the same collision cell (MPC).

In order to gain a deeper understanding of the relations between DPD and MPC,
we propose two new methods: DPD with a multibody thermostat (DPD-MT) and
MPC-Langevin dynamics (MPC-LD).
First, we modify the DPD thermostat $f_{\rm {DT1}}$ into a multibody 
thermostat to make DPD more similar to MPC.
The number of the thermostats in $f_{\rm {DT1}}$ is $3N N_{\rm {nb}}/2$,
which is more than the number of degrees of freedom $3N$ for 
$N_{\rm {nb}}>2$.  Since the excess thermostats do not play a role,
we consider the reduction to three thermostats per particle.
We define the thermal force in DPD-MT as
\begin{equation}\label{eq:dpdmt}
 f_{\rm {DT2}} =- w_i^{\rm 0} ({\bf v}_{i}-{\bf v}_{i}^{\rm G} )
 + \sqrt{w_i^{\rm 0}}{\bf \xi}_{i}(t)
+ \sum_{j\not=i} w(r_{ij})\left\{({\bf v}_{j}-{\bf v}_{j}^{\rm G} ) 
- \frac{{\bf \xi}_{j}(t)}{\sqrt{w_j^{\rm 0}}}\right\},
\end{equation} 
where $w_i^{\rm 0}=\sum_{j\not=i} w(r_{ij})$, and  
${\bf v}_{i}^{\rm G}= \sum_{j\not=i} w(r_{ij}){\bf v}_{j}/w_i^{\rm 0}$ is
the weighted mean velocity. 
The first term of $f_{\rm {DT2}}$ is the friction term of $f_{\rm {DT1}}$, and
$N_{\rm {nb}}/2$ thermostats are unified into one thermostat
between $i$-th particle and its neighbors.
The third and fourth terms are needed to conserve the translational momentum.
The Fokker-Planck equation for DPD-MT is found to be
\begin{eqnarray}\label{eq:dpdmt2}
&& \frac{\partial P({\bf X},t)}{\partial t} = \sum_{i} 
\left\{ -{\bf v}_i\cdot\nabla_i + \frac{(\nabla_i U)}{m}\cdot\partial_i 
+ \partial_i T_i \right\} P({\bf X},t), \nonumber \\
&&{\hspace{-12pt}} T_i =  w_i^{\rm 0} ({\bf v}_i -2{\bf v}_i^{\rm G}  
+ \frac{k_{\rm B}T}{m} \partial_i )
{\hspace{6pt}} +\sum_{j\not=i} w(r_{ij}) \Biggr[ {\bf v}_j^{\rm G} 
+ \frac{k_{\rm B}T}{m} \biggr\{ 
-2\partial_j + \sum_{k\not=j}\frac{w(r_{jk})}{ w_j^{\rm 0}} \partial_k \biggl\} \Biggl],
\end{eqnarray}
where ${\bf X}=\{ ({\bf r}_i,{\bf v}_i)|i=1,..,N \}$ and 
$\partial_i= \partial/\partial {\bf v}_i$.  The steady state 
$\partial P({\bf X},t)/\partial t=0$ is obtained in thermal equilibrium.
The angular momentum is conserved (DPD-MT+a), when the thermostat for the 
$i$-th particle is applied only in the direction ${\bf r}_i-{\bf r}_i^{\rm G}$,
where the weighted center of mass is 
${\bf r}_{i}^{\rm G}= \sum_{j\not=i} w(r_{ij}){\bf r}_{j}/w_i^{\rm 0}$.
We checked that Shardlow's S1 splitting algorithm~\cite{shar03} can be applied to DPD-MT.

Next, we modify the MPC method to the Langevin version (MPC-LD), with
\begin{eqnarray}\label{eq:mpld0}
 f_{\rm {MPLT}} = - \gamma ({\bf v}_i-{\bf v}_{\rm c}^{\rm G} )
 + \sqrt{\gamma}  \left\{ {\bf \xi}_i(t) - \sum_{j \in {\rm cell}} \frac{{\bf \xi}_j(t)}{N_{\rm {c}}} \right\}.
\end{eqnarray}
The thermostat is applied to the relative velocities in a collision cell.
MPC-LD+a is given by the addition of
the angular-momentum constraint in the cell,
\begin{eqnarray}\label{eq:mpldan}
 f_{\rm {AMC}} = 
\left[ m{\bf \Pi}^{-1} \sum_{j \in {\rm cell}} {\bf r}_j\times \{\gamma {\bf v}_j - \sqrt{\gamma}{\bf \xi}_j(t)\}\right]\times {\bf r}_{i}.
\end{eqnarray}
The numbers of the thermostats in MPC-LD-a ($f_{\rm {MPLT}}$) and in MPC-LD+a 
($f_{\rm {MPLT}}+f_{\rm {AMC}}$) are $3(N-N_{\rm {cell}})$ and 
$3(N-2N_{\rm {cell}})$, respectively,
where $N_{\rm {cell}}$ is the number of the cells occupied by particles.
The discrete equation for MPC-LD-a is given by the leapfrog algorithm,
\begin{eqnarray}\label{eq:mpldlf}
{\bf v}_{i}(t_{n+1}) = {\bf v}_{\rm c}^{\rm G} 
    + a\{{\bf v}_{i}(t_n)-{\bf v}_{\rm {c}}^{\rm G} \} 
    + b\{{\bf \xi}_{i,n} -\sum_{j \in {\rm cell}} 
                   \frac{{\bf \xi}_{j,n}}{N_{\rm {c}}}\},
\end{eqnarray} 
where $a$ and $b$ are given by Eq.~(\ref{eq:lgvab}).
Eq.~(\ref{eq:mpldlf})  with $\gamma \Delta t/2m=1$  corresponds to 
Eq.~(\ref{eq:mpcat}) of MPC-AT.
Energy conservation can be added  by a rescaling of 
the relative velocities in each cell.
Eq.~(\ref{eq:mpldlf}) resembles Eq.~(\ref{eq:mpc0}) of MPC.
The correlation $\langle {\bf v}_{i}(t_{n+1}){\bf v}_{i}(t_n) \rangle$ 
decreases 
with increasing $\gamma \Delta t/2m$ (MPC-LD) or angle $\phi$ (MPC).
We checked that 
the correct (flat) radial distribution function of the ideal gas
is obtained
by all MPC and DPD methods with the S1 splitting algorithm~\cite{shar03}, 
unlike for some other DPD integrators such as the modified 
velocity-Verlet algorithm in Ref.~\cite{groo97}.
When only two particles are in the cell,
the MPC-LD thermostat corresponds the pairwise thermostat in DPD.
The Langevin versions of DPD and MPC have the same relation as the 
Anderson-thermostat versions.

The difference between DPD-MT and MPC-LD is the way in which neighboring particles 
are selected.  DPD-MT reduces the number of DPD thermostats.
However, it does not reduce the numerical costs of simulations,
since the vectors ${\bf r}_{ij}$ of all neighboring particles have to 
be calculated.  In MPC-LD-a, the numerical costs are reduced;
the only information about particle positions needed is their 
partitioning into cells.  When a very small time step is chosen, 
$\Delta t\ll l_{\rm c}\sqrt{m/k_{\rm B}T}$, particles only move a 
small distance compared to the cell size in $\Delta t$; in this case,
the random-shift procedure of collision cells at each time 
step implies that the particles can only react to the time average of 
the MPC-LD thermostats. This average gives an effective weight
$w_{\rm {sq}}({\bf r}_{ij})=|(1- x_{ij}/l_{\rm c})(1- y_{ij}/
      l_{\rm c})(1- z_{ij}/l_{\rm c})|$ for $|\alpha_{ij}| < l_{\rm c}$
and $w_{\rm {sq}}({\bf r}_{ij})=0$ otherwise, where $\alpha \in \{x,y,z\}$.
This is similar to DPD-MT with the weight function $w_{\rm {sq}}({\bf r}_{ij})$,
where the rotational symmetry is broken because of $w_{\rm {sq}}({\bf r}_{ij})$.
Thus, MPC-LD can be interpreted as a version of DPD-MT, which 
approximates the weight $w_{\rm {sq}}({\bf r}_{ij})$ by 
an uniform weight inside the randomly shifted cell.

\begin{figure}
\onefigure{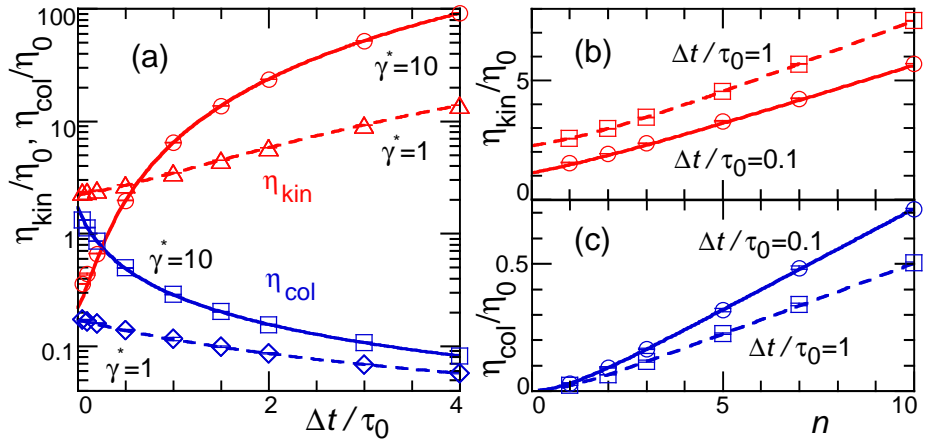}
\caption{  \label{fig:vis}
(Color online)
Viscosity dependence of MPC-LD-a on (a) the time step $\Delta t$ at $n=3$
and (b), (c) the mean number $n$ of particles per cell at $\gamma^*=1$.
Symbols represent simulation data, 
lines indicate the results of Eqs.~(\ref{eq:kinv}) and (\ref{eq:colv}).
Viscosity and time are given in units of 
$\eta_{\rm 0}=\sqrt{mk_{\rm B}T}/{l_{\rm c}}^2$ and 
$\tau_{\rm 0}=l_{\rm c}\sqrt{m/k_{\rm B}T}$, respectively.
The reduced friction constant is $\gamma^*=\gamma\tau_{\rm 0}/m$.
Error bars are estimated from three independent runs and are much smaller
than the size of the symbols.
}
\end{figure}

Although we have introduced DPD-MT and MPC-LD mainly to fill the missing 
links in Fig.~\ref{fig:rela}, they can be used for practical applications.
MPC-LD and MPC-AT need less computational costs than DPD for high densities,
and have stronger thermostats than MPC.
As an example, we investigate here the viscosity of MPC-LD-a with the
ideal-gas equation of state.
The viscosities of other methods will be reported elsewhere.
The viscosity consists of two contribution;
the kinetic viscosity $\eta_{\rm {kin}}$ and the 
collision viscosity $\eta_{\rm {col}}$ result from the momentum transfer
due to particle displacements and collisions, respectively.
The theoretical derivation of the viscosity for MPC~{\cite{kiku03}}
can be straightforwardly applied to MPC-LD.
The viscosities are then obtained as 
\begin{eqnarray} 
\label{eq:kinv}
\eta_{\rm {kin}} &=& \frac{n k_{\rm B}T }{ {l_{\rm c}}^d }
\left[ \frac{n(1+\gamma\Delta t/2m)^2}{(2\gamma/m) (n-1+e^{-n}) } 
          - \frac{\Delta t}{2} \right], \\
\eta_{\rm {col}} &=& \frac{\gamma (n-1+e^{-n}) }{ 12 {l_{\rm c}}^{d-2}
                    (1+ \gamma\Delta t/2m) }
\label{eq:colv}
\end{eqnarray}
where $n=\langle N_{\rm c}\rangle$ and $d$ is the spatial dimension.
We also calculate the viscosity from simulations for simple shear flow
with Lees-Edwards boundary conditions~\cite{kiku03,alle87} in 
three dimensions.  Fig.~{\ref{fig:vis}} shows that the theoretical and 
numerical results are in very good agreement.
As $\Delta t$ or $\gamma/m$ increases, 
$\eta_{\rm {kin}}$ increases and $\eta_{\rm {col}}$ decreases.
Thus, the viscosity of MPC-LD can be varied easily.

In summary,
we have proposed new mesoscale simulation techniques --- DPD-MT, MPC-LD, 
and variations of MPC-AT --- and clarified the relations between 
several particle-based hydrodynamic methods.
An obvious question is now which of these methods should be used for a 
given application. The answer  
depends on the system under investigation and the computational demands.
For example, the Reynolds number and other dimensionless hydrodynamic 
quantities typically have to be adjusted to match experimental conditions.
In general, the methods of the MPC group reduce computational costs   
compared to the DPD group, but have the disadvantage of (weakly) 
breaking the rotational symmetry.
We have demonstrated here that a comparison of different simulation 
techniques can stimulate the development of new methods,
and that ideas developed for one technique can be employed fruitfully 
in other techniques.

\acknowledgments
We thank M.~Ripoll for helpful discussions. 
G.G.~acknowledges partial support of this work by the DFG 
through the priority program ``Nano- and Microfluidics''.

\end{document}